\documentclass[fleqn,twoside,twocolumn,nofootinbib]{revtex4} 
\usepackage{ujp} 

\begin{document}
\title[Quantum-thermal fluctuations of effective macroparameters]
{Quantum-thermal fluctuations of effective macroparameters and their correlations}%
\author{A.D.~Sukhanov}
\affiliation{N.N.~Bogoliubov Laboratory of theoretical physics, JINR}
\address{Dubna, Russia}
\email{ogol@mail.ru}
\author{O.N.~Golubjeva}
\affiliation{Peoples' Friendship University of Russia}
\address{Moscow, Russia}
\email{ogol@oldi.ru}
\author{V.G.~Baryakhtar}
\affiliation{Institute of Magnetism, Nat. Acad. of Sci. of Ukraine}
\address{Kyiv, Ukraine}
\email{bvg@mail.vtv.kiev.ua}
\udk{???}  \razd{\secix}

\newcommand{\E}{\mathcal{E}}
\newcommand{\J}{\mathbb{J}}
\newcommand{\D}{\partial}
\newcommand{\A}{\mathcal{E}_{ef}}
\newcommand{\T}{\mathbb T}
\newcommand{\B}{\mathbb S}
\newcommand{\Ch}{\mathop{\rm ch}}
\newcommand{\Sh}{\mathop{\rm sh}}

\setcounter{page}{1120}%
\maketitle

\begin{abstract}
The application of the conventional theory of macroparameter
fluctuations has been shown to go beyond the framework of
the thermodynamic description in a number of cases. The principles of
the theory of quantum-thermal fluctuations of effective
macroparameters and their correlations have been formulated. The
theory satisfies the applicability conditions of equilibrium
thermodynamics and is based on effective macroparameters, which take
the integral stochastic action of the environment into account at
any temperatures. The correlator of conjugate macroparameters,
namely, the effective entropy and the effective temperature, has
been calculated. The correlator was found to be proportional to the
effective action which characterizes the stochastic environment. The
pair correlators for the conjugate effective parameters
\textquotedblleft entropy--temperature\textquotedblright\ and
\textquotedblleft coordinate--momentum\textquotedblright\ have been
demonstrated to depend linearly on the effective action, with their
minimum values being determined by Planck's constant.
\end{abstract}

\section{Problem of Accounting for Quantum Effects in the Macroparameter
Fluctuation Theory}

In the last years, the necessity in the application of thermodynamics to
relatively small objects (nanoparticles, nuclear spins, and so forth), which
are in the thermal equilibrium at low temperatures, has been recognized. Thus,
a necessity to attentively analyze the fundamentals of thermodynamics arose.

The equilibrium thermodynamics is known to be based upon four principles
(laws). Among them, the zeroth law, which introduces the fundamental idea of
thermal equilibrium between an object and its macroenvironment (called a
thermostat), is primordial. In the \textit{classical} thermodynamics, all
macroparameters are determined rigorously, so that the zeroth law looks like a
condition for the \textquotedblleft strict equality\textquotedblright\ between the
temperatures of the object, $T$, and the thermostat, $T_{0}$,
\begin{equation}
T=T_{0}.\label{1}%
\end{equation}
This relation is equivalent to understanding the temperature, which is
measured on the Kelvin scale, as a conditional \textquotedblleft
marker\textquotedblright\ for the thermal equilibrium.

In \textit{statistical} thermodynamics \cite{1,2,3,4,5}, including the
macroparameter fluctuation theory, every macroparameter $A$ of the object,
including the temperature, is regarded as a random variable characterized by
fluctuations $\delta A$ around its average value $\left\langle A\right\rangle
$. To preserve the terminology used in the traditional thermodynamic
description, the additional requirement is put forward simultaneously, namely,
the relative dispersion of an arbitrary macroparameter $A$ is confined by the
condition\footnote{In the cases where the denominator in Eq.~(\ref{2}) is
nullified, Bogoliubov's quasiaverage \cite{6} rather than Gibbs's average has
to be used.}
\begin{equation}
\frac{D(A)}{\langle A\rangle^{2}}\leqslant1.\label{2}%
\end{equation}
Here,
\[
D(A)\equiv\langle(\delta A)^{2}\rangle=\langle A^{2}\rangle-\langle
A\rangle^{2}%
\]
is the dispersion of macroparameter $A$ calculated by averaging over a
distribution typical of the macroparameter fluctuation theory \cite{1,2,3}.
Below, for convenience, we use the standard deviation $\Delta A=\sqrt{D(A)}$.

In this case, the concept of thermal equilibrium obtains a
generalized meaning.  It is understood that the object temperature
can also fluctuate (a \textquotedblleft soft\textquotedblright\
equilibrium condition) owing to the thermal stochastic action by the
thermostat, which is characterized by the Boltzmann constant $k_{\rm
B}$. This means that a requirement similar to inequality (\ref{2})
is imposed on the magnitude of $\Delta T$ as well. At the same time,
the temperature of the thermostat, as a system with the infinite
number of degrees of freedom, does not fluctuate, i.e. $\Delta
T_{0}=0$.

As a result, the zeroth law in the statistical thermodynamics is expressed as
a collection of two conditions,
\begin{equation}
T=T_{0}\pm\Delta T;\;\;\;\;\frac{(\Delta T)^{2}}{T_{0}^{2}}\leqslant
1,\label{3}%
\end{equation}
where $\Delta T$ is the standard deviation from $\langle T\rangle $,
and $(\Delta T)^2$, according to the notations in formula (\ref{2}),
is the temperature dispersion. Hence, only the average temperature
of the object, $\langle T\rangle$, coincides now with the
temperature of thermostat, $T_{0}$.

As is known, in the standard fluctuation theory \cite{2,3},
expressions for dispersions of arbitrary macroparameters can be
obtained on the basis of the distribution with the modulus $\Theta$.
Involving phenomenological results, the quantity $\Theta$ is usually
written down in the form $\Theta=k_{\rm B}T_{0}$. We attract
attention to the fact that this expression implies that a model
consisting of a collection of classical oscillators is implicitly used for
the thermostat. Below, we refer to it as the classical model.

In particular, A. Einstein also found formulas \cite{1} for the temperature
dispersion,
\begin{equation}
(\Delta T)^{2}=\frac{1}{k_{\rm B}C_{V}}\Theta^{2}=\frac{k_{\rm
B}}{C_{V}}T_{0}^{2},
\label{4}%
\end{equation}
and the dispersions of the object's internal energy $U$ at a constant volume
$V$,
\begin{equation}
(\Delta U)^{2}=\frac{C_{V}}{k_{\rm B}}\Theta^{2}=k_{\rm B}C_{V}T_{0}^{2}, \label{5}%
\end{equation}
where
\[
C_{V}(T,V)=\left.  \frac{\partial U}{\partial T}\right\vert _{V}%
\]
is the heat capacity of the object.

It is of interest to elucidate which are the objects, for which the expression
obtained above satisfies the inequality in condition (\ref{2}). The answer is
contained in a specific expression for the heat capacity. For instance, for
the macroscopic objects consisting of atoms, $U\sim N$ and $(C_{V})\sim N$, where
$N$ has an order of the Avogadro number. Hence, the dispersion $(\Delta T)^{2}%
\sim\frac{1}{N}$. Accordingly, $(\Delta U)^{2}\sim N$. Therefore, condition
(\ref{2}) is satisfied for the relative dispersions of internal energy and
temperature in such objects at high temperatures, if $N\gg1$.

At the same time, we attract attention to the fact that, for
example, in the case of a single classical oscillator ($N=1$), for
which $U=k_{\rm B}T$ and $C_{V}=k_{\rm B}$, and which is in
equilibrium with the thermostat, we have
\begin{equation}
 \frac{(\Delta U)^{2}}{\langle U\rangle^{2}}=1;\quad\frac{(\Delta T)^{2}}{\langle
T\rangle^{2}}=1.\nonumber
\end{equation}
Thus, the condition of macroscopicity ($N\gg1$) is not strictly mandatory,
while carrying out calculations in the framework of classical statistical
mechanics and in the high temperature range.

A different situation arises at relatively low temperatures, when
quantum-mechanical effects manifest themselves. In this case, the main attention
is still given to the heat capacity
\[
(C_{V})_{\rm qu}=\left.  \frac{\partial U_{\rm qu}}{\partial
T}\right\vert _{V},
\]
but, for its calculation, the internal energy $U_{\rm qu}$ is
applied, which is calculated now in the framework of the quantum
statistical mechanics. At the same time, the model of thermostat
does not undergo changes and remains classical. On this route,
the satisfactory results can still be obtained for systems of particles
with $N\gg1$ (except for, maybe, the region of ultralow
temperatures). However, there emerge the problems for the equilibrium
thermal radiation (where the concept of the number of particles is absent)
and a single quantum-mechanical oscillator.

Let us demonstrate the aforesaid using the quantum-mechanical oscillator as an
example. Its internal energy is equal, according to A. Einstein \cite{7}, to
\begin{equation}
U_{\rm
qu}=\frac{\hbar\omega}{\exp\{2\varkappa\frac{\omega}{T}\}-1}=\frac
{\hbar\omega}{2}\,\frac{\exp\{-\varkappa\frac{\omega}{T}\}}{\sinh
(\varkappa\frac{\omega}{T})},\label{6}%
\end{equation}
and the heat capacity is
\[
(C_{V})_{\rm qu}=k_{\rm B}\left(  \frac{\hbar\omega}{k_{\rm
B}T}\right) ^{2}\frac
{\exp\{2\varkappa\frac{\omega}{T}\}}{(\exp\{2\varkappa\frac{\omega}%
{T}\}-1)^{2}}=
\]
\begin{equation}
=k_{\rm B}\left(  \varkappa\frac{\omega}{T}\right) ^{2}\frac
{1}{\sinh^{2}(\varkappa\frac{\omega}{T})},\label{7}%
\end{equation}
where the notation
\[
\varkappa=\hbar/2k_{\rm B},
\]
was introduced. In accordance with the general formula (\ref{5}), the
dispersion of the internal energy of a quantum-mechanical oscillator
at the replacement of $(C_{V})$ by $(C_{V})_{\rm qu}$ looks like
\[
(\Delta U_{\rm qu})^2=k_{\rm B}(C_V)_{\rm qu}T^2=\left(\frac{\hbar
\omega}{2}\right)^2\frac{1}{\sinh^2(\varkappa\frac{\omega}{T})}=\]
\begin{equation}\label{8}
=\hbar\omega U_{\rm qu}+ U_{\rm qu}^2=
\exp\left\{2\varkappa\frac\omega T\right\}U_{\rm qu}^2,
\end{equation}
and the relative dispersion of its internal energy reads
\begin{equation}
\frac{(\Delta U_{\rm qu})^{2}}{U_{\rm qu}^{2}}=\frac{\hbar\omega}{U_{\rm qu}}%
+1=\exp\left\{2\varkappa\frac{\omega}{T}\right\}.\label{9}%
\end{equation}
A similar result is obtained for the relative dispersion of the
thermal radiation energy in the spectral interval $(\omega,\omega+\Delta\omega)$ in
the volume $V$,
\begin{equation}
\frac{(\Delta U_{\omega})^{2}}{U_{\omega}^{2}}=\frac{\hbar\omega}{U_{\omega}%
}+\frac{\pi^{2}c^{3}}{V\omega^{2}\Delta\omega}=\frac{\pi^{2}c^{3}}{V\omega
^{2}\Delta\omega}\exp\left\{2\varkappa\frac{\omega}{T}\right\}.\label{10}%
\end{equation}

Those facts evidently testify to the inapplicability of the selected method of
calculation; namely, the relative dispersions of internal energy (\ref{9}) and
(\ref{10}) do not satisfy condition (\ref{2}) for the thermodynamic
description to be applicable. Note that this circumstance did not draw a special
attention of researchers. A.I.~Anselm \cite{8} was probably the only who
marked it. In this connection, he pointed out that \textquotedblleft the
conventional thermodynamics is not applicable as the temperature falls
down\textquotedblright. This means that the statistical thermodynamics
considered as a macrotheory cannot be based exclusively upon the quantum
statistical mechanics as a microtheory.

In our opinion, all that originates from the fundamental credo of quantum
statistical mechanics, according to which the quantum-mechanical and thermal
actions of the environment can be taken into consideration independently. In
this connection, the standard routine consists in that the quantum-mechanical
characteristics of a system of microobjects are determined firstly (at $T_{0}%
=0$), and only then the system is embedded into a classical
thermostat with the distribution modulus $k_{\rm B}T_{0}$. At the
same time, formula (\ref{6}) for a quantum-mechanical oscillator
and an analogous formula for the internal energy of the thermal
irradiation, which are confirmed by experiment, testify that those
two types of stochastic action manifest themselves, as a rule,
jointly and nonadditively.

In order to overcome the arisen problems, we developed a theory
\cite{8,9,10,11}, which is based on a combined account of the
quantum-mechanical and thermal stochastic actions by the environment. The
theory denies the classical model of thermostat in favor of the
quantum-mechanical one, including the case of heat capacity calculations.
Moreover, in this case, the introduction of a different \textquotedblleft
marker\textquotedblright\ for the thermal equilibrium state is required. The
matter is that the Kelvin temperature used for such purposes in the quantum
statistical mechanics turns out noninformative in the region, where the
quantum-mechanical and thermal effects jointly manifest themselves, because
this characteristic reflects only the thermal stochastic action (in terms of
the Boltzmann constant). In addition, it is adopted to be equal to zero in
quantum mechanics, where, nevertheless, the stochastic action (although being
already of the quantum-mechanical type in terms of Planck's constant)
takes place by essence.

Purely intuitive considerations stimulate us to adopt the expression
that follows from the Planck formula for the energy of
quantum-mechanical oscillator $U_{\rm Pl}$, as a new
\textquotedblleft marker\textquotedblright; namely,
\begin{equation}
\mathbb{T}\equiv\frac{U_{\rm Pl}}{k_{\rm
B}}=\frac{\hbar\omega}{2k_{\rm B}}\coth
\frac{\hbar\omega}{2k_{\rm B}T}=\varkappa\omega\coth\frac{\varkappa\omega}%
{T},\label{11}%
\end{equation}
where the notation $\varkappa=\hbar/2k_{\rm B}$ introduced earlier
is used. We define this quantity as the effective temperature. It is
important that, at the Kelvin temperature $T=0$, the effective
temperature has a nonzero minimum,
\begin{equation}
\mathbb{T}\,^{\min}=\varkappa\omega\neq0.\label{12}%
\end{equation}

In the general case, according to formulas (\ref{11}) and
(\ref{12}), the effective temperature is a two-parameter function
$\mathbb{T}=f(T,\omega)$. It can be reduced to a one-parameter
function only in the limiting cases $T=0$ and
$T\gg\mathbb{T}\,^{\min}$, i.e., where
$\mathbb{T}\approx\mathbb{T}\,^{\min}$ and $\mathbb{T}\approx T$,
respectively. Its simultaneous dependence on the Planck and
Boltzmann constants allows one to assert that, in view
of the relation $\varkappa=\frac{\hbar}{2k_{\rm B}}$, the stochastic
action of the \textquotedblleft thermal\textquotedblright\ type
cannot be ignored even at the Kelvin temperature $T_{0}=0$.

On this basis, we constructed a theory of
fluctuations for macroparameters and their correlations, which is presented below. In so doing and
keeping preserved all traditional relations between thermodynamic quantities
coupled with the temperature, we introduced the corresponding effective
macroparameters, which are the same functions of the effective temperature.

\section{Effective Temperature and Effective Internal Energy Fluctuations}

To calculate the dispersions of macroparameters taking the
stochastic action of the quantum-mechanical type into account at any
temperatures, let us use the results of work \cite{9}, which is a
macrodescription in the framework of a theory called by us the
modern stochastic thermodynamics \cite{11}. It is based on the Gibbs
distribution in the space of macroparameters \cite{12,5}.

To describe the stochastic environment, we introduce a
quantum-mechanical model (a quantum thermostat) that combines the concepts
of vacuum and thermostat. It comprises an infinite set of quantum-mechanical
normal modes of all frequencies at any fixed Kelvin temperature, being an
analog of equilibrium thermal radiation. The effective temperature of the
object (\ref{11}) is supposed to fluctuate as well with a characteristic
frequency $\omega$, so that the zeroth law reads
\begin{equation}
\mathbb{T}=\mathbb{T}_{0}\pm\Delta\mathbb{T } \label{13}%
\end{equation}
in the developed theory, where $\mathbb{T}_{0}$ is the effective temperature
of the thermostat, and $\Delta\mathbb{T}$ is the standard deviation of the
effective temperature of the object under the equilibrium conditions.

From the modern viewpoint, the primordial principle of the
fluctuation theory of macroparameters is the principle of entropy maximum in the thermal
equilibrium state. The exponential Gibbs form for the distribution in the
space of macroparameters follows from it \cite{12},
\begin{equation}
dW(\mathcal{E})=\rho(\mathcal{E})d\mathcal{E}=\frac{1}{\Theta}\exp
\{-\frac{\mathcal{E}}{\Theta}\}d\mathcal{E}.\label{14}%
\end{equation}
Here, $\mathcal{E}$ is a random value of the internal energy of a macroobject,
which depends on the macroparameters that characterize the type of its
interaction with the environment.

One should bear in mind that, in the initial version of the
macroparameter fluctuation theory, the distribution modulus
corresponded to the classical model of thermostat with the Kelvin
temperature $T_{0}$. Since the developed theory uses the
quantum-mechanical model of thermostat, the corresponding
distribution modulus is expressed in terms of the effective
thermostat temperature $\mathbb{T}_{0}$ like $k_{\rm
B}\mathbb{T}_{0}$. As a result, the standard formula for the
macroparameter fluctuation distribution \cite{1,2,3} and the
expressions for the dispersions of any macroparameters, which follow
from it, remain formally invariable, but they are expressed now in
terms of effective macroparameters. Therefore, the dispersion of the
effective temperature of a macroobject, instead of formula (\ref{4}),
looks like
\begin{equation}
(\Delta\mathbb{T})^{2}=\frac{k_{\rm B}}{\mathbb{C}_{V}}\mathbb{T}_{0}%
^{2},\label{15}%
\end{equation}
\begin{equation}
\mathbb{C}_{V}\equiv\frac{\partial\mathbb{U}}{\partial\mathbb{T}}\label{16}%
\end{equation}
is the effective heat capacity of the object. Accordingly, for the dispersion
of the effective internal energy, instead of formula (\ref{5}), we obtain
\begin{equation}
(\Delta\mathbb{U})^{2}=k_{\rm B}\mathbb{C}_{V}\mathbb{T}_{0}^{2}.\label{17}%
\end{equation}
In other words, all expressions (\ref{13})--(\ref{17}) preserve their earlier
forms in the new theory. In this way, the concept of the equilibrium state
between an object and a stochastic environment remains also invariable.
However, this equilibrium is now a generalized concept, which involves
the existence of two nonadditive types of stochastic action in the
integral concept of effective temperature $\mathbb{T}$.

Now, let us apply the results obtained to the study of a macroobject that
can be represented as a quantum-mechanical oscillator\footnote{Note
that the scope of problems, which can be solved with the use of this
model, is wide enough, because the potential energy can be
approximated by a parabola in a vicinity of its minimum.}. For
this object, $\mathbb{U}\equiv U_{\rm Pl},$ and the effective heat
capacity $\mathbb{C}_{V}=k_{\rm B}$. Then formula (\ref{15}) for
the effective temperature dispersion looks like
\[
(\Delta\mathbb{T})^{2}=(\mathbb{T}_{0})^{2},
\]
and the relative dispersion of the effective temperature of this object
obeys condition (\ref{3}) now. It is the difference of this quantity
from the dispersion of the Kelvin temperature of the same object in
the quantum statistical mechanics. Accordingly, the relative
dispersion of its internal energy, with regard for the general
formula (\ref{17}) and the formulas $C_{V}=k_{\rm B}$ and
$\mathbb{U}=k_{\rm B}\mathbb{T}_{0},$ reads
\begin{equation}
\frac{(\Delta\mathbb{U})^{2}}{\mathbb{U}^{2}}=\frac{k_{\rm B}C_{V}(\mathbb{T}%
_{0})^{2}}{(k_{\rm B}\mathbb{T}_{0})^{2}}=1,\label{18}%
\end{equation}
and condition (\ref{2}) is also satisfied.

To make a more detailed comparison between the formulas obtained here and
those known from the quantum statistical mechanics, let us express the
dispersion of the effective internal energy of a quantum-mechanical oscillator in
the form
\begin{equation}
(\Delta\mathbb{U})^{2}=\left(  \frac{\hbar\omega}{2}\right)  ^{2}
\left[1+\sinh^{-2}\left(\varkappa\frac{\omega}{T_{0}}\right)\right].\label{19}%
\end{equation}
By comparing formulas (\ref{19}) and (\ref{8}), where the heat capacity looks
like expression (\ref{7}), we can present the second term in Eq.~(\ref{19}) in the
form that reminds the initial formula (\ref{5}), but includes an explicit
dependence on the Kelvin temperature $T_{0}$,
\begin{equation}
(\Delta\mathbb{U})^{2}=\left(\frac{\hbar\omega}{2}\right)^{2}+k_{\rm B}(C_{V})_{\rm qu}T_{0}%
^{2}.\label{20}%
\end{equation}
Expression (\ref{20}) evidently differs from expression (\ref{9}) by an
additional term, which can be written down in the form
\begin{equation}
\left(\frac{\hbar\omega}{2}\right)^{2}=\frac{\hbar}{2}\rho_{\omega}(\omega,0)\omega
^{2},\label{21}%
\end{equation}
\begin{equation}
\rho_{\omega}(\omega,0)\equiv\left.  \frac{\partial\mathbb{U}}{\partial\omega
}\right\vert _{T=0}=\frac{\hbar}{2}\label{22}%
\end{equation}
is the spectral density of the effective internal energy at $T_{0}=0$. Then,
formula (\ref{20}) reads
\begin{equation}
(\Delta\mathbb{U})^{2}=\frac{\hbar}{2}\rho_{\omega}(\omega,0)\omega^{2}%
+k_{\rm B}(C_{V})_{\rm qu}T_{0}^{2}.\label{23}%
\end{equation}

It is worth to note that, in contrast to the quantum-statistical formula
(\ref{8}) valid for quantum-mechanical oscillators, formula (\ref{23}) of
the developed theory contains an additional term, which reveals itself at
$T_{0}=0$ as well. Really, in the limit $T_{0}\rightarrow0$, the second
term in formula (\ref{23}) disappears, so that
\begin{equation}
(\Delta\mathbb{U}\,^{\min})^{2}\!=\frac{\hbar}{2}\rho_{\omega}(\omega,0)\omega
^{2}=(\mathbb{U}\,^{\min})^{2}=\left(\!\frac{\hbar\omega}{2}\!\right)^{2}\!\neq0.\label{24}%
\end{equation}
In the quantum statistical mechanics, on the contrary, $(\Delta U_{\rm qu}%
)^{2}\rightarrow0$ at $T_{0}\rightarrow0$. Hence, in the theory developed
here, the validity of condition (\ref{2}) for the effective internal energy
both at an arbitrary Kelvin temperature and at $T_{0}\rightarrow0$ is
essentially associated with the account for the energy of zero oscillations.

In our opinion, another important result was also obtained in this consideration. It consists in
that the zero energy has not also an exact invariable value, but possesses
a certain ``dispersion'' or ``width'', by
fluctuating within its limits. Naturally, there arises a question:
\textquotedblleft At the expense of what does it take place?\textquotedblright%
\ The answer is that, in this case, the system is in the equilibrium contact
with \textquotedblleft cold\textquotedblright\ (in the Kelvin sense) vacuum,
which renders a quantum-mechanical stochastic action. The latter results in the
randomization of the internal energy of an object even at the absolute zero temperature.

\section{Fluctuations of Effective Entropy and Their Correlation with
Fluctuations of Effective Temperature}

Under the thermal equilibrium conditions, besides the fluctuations of
individual macroparameters, $\delta A$ and $\delta B$, a considerable role is
also played by the correlation between them. The measure of this correlation
is known to be given by the expression $\sigma_{AB}\equiv\langle\delta A,\delta
B\rangle$ referred to in the general case as the correlator.

Let us analyze the correlation between the fluctuations of effective
macroparameters. In so doing, we are interested in a nontrivial case of
conjugate effective macroparameters. It is known that the concept of conjugate
quantities is one of the key ones in quantum mechanics. Nevertheless, it is
also used in thermodynamics, but, as a rule, on the basis of heuristic
considerations. While analyzing the thermodynamic potentials (see, e.g., work
\cite{13} by Sommerfeld), it hits you in the eye that they contain some stable
combinations of macroparameters, such as $(Ada)$ or $(adA)$. For instance, the
temperature is always combined with the entropy, and the pressure with the
volume. Physically, those relations reveal themselves in every pair in the
existence of a nonzero correlator between macroparameter fluctuations, which
reflects their mutual interdependence. This circumstance forms the initial
base to consider the pair of effective quantities $\mathbb{S}$ and
$\mathbb{T}$ as thermodynamically conjugate, like the pair
coordinate--momentum in quantum mechanics.

To substantiate this statement within the developed theory, let us
calculate the dispersion of the effective entropy and the correlator between the
fluctuations of the effective entropy and the effective temperature. Taking
into account that, at a fixed effective volume $\mathbb{V}$, the fluctuation
of the effective entropy is
\begin{equation}
\delta\mathbb{S}=\left.
\frac{\delta\mathbb{U}}{\mathbb{T}_{0}}\right\vert
_{\mathbb{V}}=\frac{1}{\mathbb{T}_{0}}\,\mathbb{C}_{\mathbb{V}}\cdot
\delta\mathbb{T},\label{25}
\end{equation}
we obtain the following expression for the corresponding dispersion:
\begin{equation}
(\Delta\mathbb{S})^{2}\equiv\langle(\delta\mathbb{S})^{2}\rangle
=\frac{\mathbb{C}_{V}^2}{\mathbb{T}_{0}^{2}}\;\langle(\delta
\mathbb{T})^{2}\,\rangle=\frac{\mathbb{C}_{V}^2}{\mathbb{T}_{0}^{2}%
}\;(\Delta\mathbb{T})^{2}. \label{26}%
\end{equation}
Taking into consideration the consequence of general formula (\ref{15}) for
the standard deviation of the effective temperature, $(\Delta\mathbb{T}%
)=(k_{\rm
B})_{0}^{1/\,2}\mathbb{T}(\mathbb{C}_{\mathbb{V}})^{-1/\,2}$, and
extracting the square root of expression (\ref{26}), we obtain that
the standard deviation of the effective entropy reads
\begin{equation}
\Delta\mathbb{S}=\frac{\mathbb{C}_{V}^2}{\mathbb{T}_{0}}\,(\Delta
\mathbb{T})=\big(k_{\rm B}\mathbb{C}_{V}^2\big)^{1/2}. \label{27}%
\end{equation}

To derive the required fluctuation correlator $\sigma_{\mathbb{ST}}$, let us
use formula (\ref{25}) once more. We obtain that this correlator is
proportional to the temperature dispersion:
\begin{equation}
\sigma_{\mathbb{S}\mathbb{T}}=\langle\delta\mathbb{S}\cdot\delta
\mathbb{T}\rangle=\frac{\mathbb{C}_{V}}{\mathbb{T}_{0}}\;\langle
\delta\mathbb{T}\cdot\delta\mathbb{T}\rangle=\frac{\mathbb{C}_{V}}%
{\mathbb{T}_{0}}(\Delta\mathbb{T})^{2}.\label{28}%
\end{equation}
Taking Eq.~(\ref{15}) into account, the corresponding fluctuation correlator
is
\begin{equation}
\sigma_{\mathbb{S}\mathbb{T}}=\frac{\mathbb{C}_{V}}{\mathbb{T}_{0}}%
\;\frac{k_{\rm B}}{\mathbb{C}_{V}}(\mathbb{T}_{0})^{2}=k_{\rm B}\mathbb{T}%
_{0},\label{29}%
\end{equation}
so that the dependence on the effective heat capacity disappears. In other
words, the correlator of the effective macroparameters $\sigma_{\mathbb{ST}}$ is
governed only by a parameter of the quantum thermostat, namely, by its effective
temperature, which, in accordance with Eq.~(\ref{12}), does not equal zero in
principle. Therefore, it behaves like a quantum-mechanical correlator between
the variables $p$ and $q$ in quantum mechanics, which also does not equal
zero. This circumstance comprises an additional argument to consider the
effective macroparameters $\mathbb{S}$ and $\mathbb{T}$ as conjugate quantities.

\section{Interrelation between the Correlation of Conjugate Micro- and
Macroparameter Fluctuations and the Effective Action
by~the~Environment}

To elucidate the physical sense of the expression for $\sigma_{\mathbb{ST}}$
in the form (\ref{29}), we address to the $(\hbar,k)$-dynamics \cite{8,10}. In so doing,
we proceed from Bogoliubov's idea \cite{14}, according to which only the
stochastic action of the environment can invoke a nontrivial correlation between fluctuations of
the micro- and macroparameters. In the framework of the $(\hbar
,k)$-dynamics \cite{10}, it was found that, at the microlevel, this action is
described by a specific operator, \textit{Schr\"{o}dingerian, }%
\begin{equation}
\hat{\jmath}\equiv\delta\hat{p}\cdot\delta\hat{q}=\hat{\sigma}-i\hat{\jmath
}_{0}, \label{30}%
\end{equation}
where
\[
\delta\hat{p}=\hat{p}-\langle|\hat{p}|\rangle;\quad\delta\hat{q}%
=\hat{q}-\langle|\hat{q}|\rangle;
\]%
\[
\hat{\sigma}\equiv\frac{1}{2}\{\delta\hat{p},\delta\hat{q}\};\quad\hat
{\jmath}_{0}\equiv\frac{i}{2}[\hat{p},\hat{q}]=\frac{\hbar}{2}\hat{I},
\]
and $\hat{I}$ is the unit operator.

As a natural measure for the stochastic action by the environment at the
microlevel, we introduced a specific parameter, the action $\mathcal{J}$,
defined as the absolute value of the averaged Schr\"{o}dingerian,
\begin{equation}
\mathcal{J}\equiv\sqrt{\mathit{\Sigma}^{2}+\frac{\hbar^{2}}{4}},\label{31}%
\end{equation}
where $\mathit{\Sigma}$ and $\frac{\hbar}{2}$ are the average values for the
operators $\hat{\sigma}$ and $\hat{\jmath}_{0}$, respectively. The quantity
\[
\mathit{\Sigma}=\big|\langle\psi^{\ast}(q)|\,\hat{\sigma}\,|\,\psi
(q)\rangle\big|=\frac{\hbar}{2}\alpha
\]
is determined by the phase of the complex wave function
\[
\psi(q)=\left[  2\pi(\Delta q)^{2}\right]  ^{-1/4}\exp\left\{  -\frac{q^{2}%
}{4(\Delta q)^{2}}(1-i\alpha)\right\}.
\]
A similar form has, e.g., the wave function of a free microparticle, which
describes a smearing of its initial state in time at the zero Kelvin
temperature \cite{15}.

However, the meaning of $\mathcal{J}$ can be interpreted differently. It can
also be regarded as the absolute value of \textquotedblleft quantum-mechanical
correlator\textquotedblright\ $\left\vert \langle\delta p|\delta
q\rangle\right\vert $ for the fluctuations of canonically conjugate
quantities, the coordinate and the momentum,
\begin{equation}
\mathcal{J}\equiv\big|\langle\,|\,\hat{\jmath}\,|\,\rangle\big|=\big|\langle
\,|\,\delta\hat{p}\cdot\delta\hat{q}\,|\,\rangle\big|=\big|\langle\,\delta
p\,|\,\delta q\,\rangle\big|.\label{32}%
\end{equation}

At the macrolevel and in the equilibrium state with the quantum thermostat,
the applicability region for definition (\ref{32}) becomes wider
\cite{8,10,11}. The role of the wave function used for averaging the
operator $\hat{\jmath}$ is played by a complex-valued wave function of warm
vacuum in the coordinate representation, which depends on the Kelvin
temperature of the quantum thermostat,
\begin{equation}
\psi_{\scriptscriptstyle{T_{0}}}(q)=\left[  2\pi(\ \Delta\mathbb{Q}%
)^{2}\right]  ^{-1/4}\exp\left\{  -\frac{q^{2}}{4(\Delta\mathbb{Q})^{2}%
}(1-i\alpha_{\scriptscriptstyle{T_{0}}})\right\},  \label{33}%
\end{equation}
where the parameter%
\begin{equation}
\alpha_{\scriptscriptstyle{T_{0}}}=\frac{1}{\sinh(\varkappa\frac{\omega}%
{T_{0}})}\label{34}%
\end{equation}
is included into the wave-function phase, and
\begin{equation}
(\Delta\mathbb{Q})^{2}=\frac{\hbar}{2m\omega}\coth\left(\varkappa\frac{\omega
}{T_{0}}\right).\label{35}%
\end{equation}
is the coordinate dispersion.

As a result, the quantity
\[
\mathbb{J}_{0}=\big|\langle\psi_{\scriptscriptstyle{T_{0}}}^{\ast}%
(q)|\,\hat{\jmath}\,|\,\psi_{\scriptscriptstyle{T_{0}}}(q)\rangle
\big|=
\]
\begin{equation}
=\sqrt{\frac{\hbar^{2}}{4}\,\alpha_{\scriptscriptstyle{T_{0}}}^{2}%
+\frac{\hbar^{2}}{4}}=\frac{\hbar}{2}\coth\left(\varkappa\frac{\omega}{T_{0}%
}\right),\label{36}%
\end{equation}
which is equivalent to $\mathcal{J}$, appears as a natural characteristic of
the integral stochastic action at the \textit{macrolevel}. The quantity
$\mathbb{J}_{0}$ was introduced earlier empirically as a new effective
thermodynamic parameter, the \textit{effective action} of the quantum thermostat,
typical of stochastic thermodynamics \cite{9,11}. Its important feature
consists in that it is formally equivalent to other effective
macroparameters of equilibrium stochastic thermodynamics and is genetically
coupled with the microdescription through the wave-function phase depending on
$\alpha_{T_{0}}$. This circumstance can be used for the interpretation of
deeper interrelations between two levels of the description of the Nature, the
quantum-mechanical theory and the thermodynamics. In particular, when
comparing Eqs.~(\ref{34}) and (\ref{36}), one can see that
\begin{equation}
\Sigma_{\scriptscriptstyle{T_{0}}}=\big|\langle\psi_{\scriptscriptstyle{T_{0}%
}}^{\ast}(q)|\,\hat{\sigma}\,|\,\psi_{\scriptscriptstyle{T_{0}}}%
(q)\rangle\big|=\frac{\hbar}{2}\alpha_{\scriptscriptstyle{T_{0}}}.\label{37}%
\end{equation}
At the same time, using formula (\ref{7}) for the heat capacity $(C_{V})_{\rm qu}%
$, we obtain
\begin{equation}
\Sigma_{\scriptscriptstyle{T_{0}}}=\frac{T_{0}}{\omega}\sqrt{k_{\rm B}(C_{V}%
)_{\rm qu}}.\label{38}%
\end{equation}
Hence, the same quantity $\Sigma_{\scriptscriptstyle{T_0}}$
characterizing the thermal action in the equilibrium state is
strictly coupled with a nonzero \textit{phase} of the warm-vacuum
wave function at the \textit{microlevel} and, simultaneously, with a
nonzero \textit{heat capacity} at the \textit{macrolevel}.

In work \cite{8}, it was shown that, in the state of equilibrium between a
quantum-mechanical oscillator and the quantum thermostat, the
quantum-mechanical correlator \textquotedblleft
coordinate--momentum\textquotedblright\ looks like
\begin{equation}
\sigma_{\mathbb{P}\mathbb{Q}}=\frac{\hbar}{2}\coth(\varkappa\frac{\omega
}{T_{0}})=\mathbb{J}_{0},\label{39}%
\end{equation}
We emphasize that, in the given context, the quantities $p$ and $q$ acquire
the meanings of the effective macroparameters $\mathbb{P}$ and $\mathbb{Q}$ of
the object in the model of quantum-mechanical oscillator, which is in the
state of equilibrium with the environment.

Changing over to the microdescription in the limit $T_{0}\rightarrow0$,
formula (\ref{39}) can be transformed to the form
\begin{equation}
\sigma_{\mathbb{P}\mathbb{Q}}^{\min}=\mathbb{J}_{0}^{\min}=\frac{\hbar}%
{2},\label{40}%
\end{equation}
where the minimum value of $\sigma_{\mathbb{P}\mathbb{Q}}$ is determined by
Planck's constant that characterizes the quantum-mechanical stochastic action.
A comparison of formulas (\ref{40}) and (\ref{36}) suggests that, in the
general case, the effective action $\mathbb{J}_{0}$ in the equilibrium state
can be presented in the form
\begin{equation}
\mathbb{J}_{0}=\frac{1}{2}\hbar^{\ast}(\hbar,k_{\rm B})\geqslant\frac{\hbar}%
{2},\label{41}%
\end{equation}
where the quantity $\hbar^{\ast}\equiv\hbar\coth(\frac{\hbar\omega}%
{2k_{\rm B}T_{0}})$ can be naturally interpreted as a generalization
of the elementary action $\hbar$ to the case with $T_{0}\neq0$.

While comparing formulas (\ref{12}) and (\ref{36}), one can see that the
quantities $\mathbb{J}_{0}$ and $\mathbb{T}_{0}$ are proportional to each
other,
\begin{equation}
\mathbb{T}_{0}=\frac{\omega}{k_{\rm B}}\mathbb{J}_{0}.\label{42}%
\end{equation}
This fact gives grounds to use $\mathbb{J}_{0},$ rather than $\mathbb{T}_{0},$
as a \textquotedblleft marker\textquotedblright\ of the equilibrium with the
environment in all thermodynamic relations.

We now consider the correlator of fluctuations of the conjugate macroparameters, the
effective entropy and the effective temperature, typical of the macrodescription.
Using interrelation (\ref{42}), we demonstrate that the correlator
$\sigma_{\mathbb{ST}}$ of the form (\ref{29}) also depends on $\mathbb{J}_{0}$.
Really, in view of formula (\ref{42}), relation (\ref{29}) can be presented in the form
\begin{equation}
\sigma_{\mathbb{S}\mathbb{T}}=\langle\delta\,\mathbb{S},\delta\mathbb{T}%
\rangle=\omega\mathbb{J}_{0}.\label{43}%
\end{equation}
Examining the limiting (at the Kelvin temperature
$T_{0}\rightarrow0$) correlator value
$\sigma_{\mathbb{ST}}=\omega\mathbb{J}_{0}^{\mathrm{\min}}$, we see
that, in this case, it is determined only by a stochastic action
of the quantum-mechanical type,
\begin{equation}
\sigma_{\mathbb{S}\mathbb{T}}^{\min}=\omega\mathbb{J}_{0}^{\min}=\mathbb{U}%
_{0}^{\min}=\frac{\hbar\omega}{2},\label{44}%
\end{equation}
where the energy of zero oscillations stands on the right.

Hence, according to formulas (\ref{40}) and (\ref{44}), the minimum values of
fluctuation correlators for both pairs of the conjugate variables, $(\mathbb{P}%
,\mathbb{Q})$ and $(\mathbb{S},\mathbb{T})$, are determined by the same
Planck's constant. As the temperature grows, the correlators $\sigma
_{\mathbb{P}\mathbb{Q}}$ and $\sigma_{\mathbb{ST}}$ increase synchronously,
remaining proportional to each other.

\section{Schr\"{o}dinger uncertainty relations and their role in the theory of fluctuations of
micro- and macroparameters}

It is known that the physical content of the uncertainties relation
(UR), introduced for the first time by W. Heisenberg in quantum
mechanics, has been associated for many years exclusively with the
theory of measurements. However, it was found that URs take place in
other theories as well: in the equilibrium thermodynamics, theory of
Brownian motion, and so forth. In our opinion, this circumstance is
related to the stochastic action of an environment, which implicitly
generates URs in those theories. From this viewpoint, a search for a
universal relation between URs arising at zero and finite
temperatures is quite justified. The analysis of fluctuations of
conjugate micro- and macroparameters, which was carried out above,
allows URs to be interpreted in the framework of the fluctuation
theory developed here.

From the mathematical viewpoint, the most general UR proposed by
E. Schr\"{o}dinger (SUR) is an implementation of the Cauchy--Buniakowski--Schwarz
inequality in the space of corresponding quantities \cite{15},
\begin{equation}
\mathcal{UP}_{AB}\equiv\Delta A\Delta B\geqslant\sigma_{AB}.\label{45}%
\end{equation}
Here, $\mathcal{U}\mathcal{P}_{AB}$ is the product of standard
derivations (the \textquotedblleft uncertainties product
\textquotedblright). The standard deviations $\Delta A$ and $\Delta
B$ play the role of a measure for the \textquotedblleft
uncertainties\textquotedblright\ of random variables, the
macroparameters $A$ and $B$. Accordingly, $\sigma_{AB}$ is the
correlator of fluctuations of those quantities.

For conjugate micro- and macroparameters, the nonzero correlator of
fluctuations imposes a restriction on the mutual behavior of standard
deviations of those quantities. To show the fundamental character of similar
restrictions, the left- and right-hand sides in relation (\ref{45}) should be
calculated independently, both in the micro- and macro-theories.

To calculate the quantity $\mathcal{U}\mathcal{P}_{\mathbb{ST}}$, which is
typical of the macrotheory, we use formulas (\ref{27}) and (\ref{15}). As a
result, we obtain that this quantity does not depend on the effective heat
capacity and looks like
\begin{equation}
\mathcal{U}\mathcal{P}_{\mathbb{S}\mathbb{T}}\equiv(\Delta\mathbb{S}%
)(\Delta\mathbb{T})=k_{\rm B}\mathbb{T}_{0}.\label{46}%
\end{equation}

Comparing formulas (\ref{46}) and (\ref{29}), we notice that, in the
state of equilibrium between the object, which is simulated as a
quantum-mechanical oscillator, and the quantum thermostat, two
different physical quantities,
$\mathcal{U}\mathcal{P}_{\mathbb{ST}}$ and $\sigma_{\mathbb{ST}}$,
have identical values. Therefore, the SUR \textquotedblleft
effective entropy--effective temperature\textquotedblright\ becomes
an equality in this case, i.e. it is saturated.
\begin{equation}\label{47}
\mathcal U\mathcal P_{\B\T} = \sigma_{\B\T}.
\end{equation}

 If\
$\mathbb{V}\neq\mathrm{const}$ owing to the growing
$\Delta\mathbb{S}$ at a immutable $\sigma_{\mathbb{ST}}$, the
discussed SUR is transformed into an inequality, so that, in the
general case, we have
\begin{equation}
\mathcal{U}\mathcal{P}_{\mathbb{S}\mathbb{T}}\geqslant\sigma_{\mathbb{S}%
\mathbb{T}}.\label{48}%
\end{equation}
Now, let us take into account that the correlator
$\sigma_{\mathbb{ST}}$ can be presented in the form (\ref{43}). It
is equivalent to the statement that mutual restrictions on the
uncertainties $\Delta\mathbb{S}$ and $\Delta\mathbb{T}$ for an
object in the equilibrium state are dictated by the integral
stochastic action of the environment, which is characterized by the
quantity $\mathbb{J}_{0}$. Therefore, in the equilibrium state, the
SUR \textquotedblleft effective entropy--effective
temperature\textquotedblright\ takes the form
\begin{equation}
\mathcal{U}\mathcal{P}_{\mathbb{S}\mathbb{T}}=\omega\mathbb{J}_{0}=\omega
\frac{\hbar^{\ast}}{2}=U_{\rm Pl}.\label{49}%
\end{equation}
We should emphasize that the left- and right-hand sides of this SUR
were obtained in the framework of the \textit{macrotheory,} in which
the expression for the Planck energy $U_{\rm Pl}$ is taken from the
experiment. In the limit $T_{0}\rightarrow0$, relation (\ref{49})
looks like
\begin{equation}
\mathcal{U}\mathcal{P}_{\mathbb{S}\mathbb{T}}^{\min}=\omega\mathbb{J}_{0}%
^{\min}=U_{\rm Pl}^{\min}=\frac{\hbar\omega}{2},\label{50}%
\end{equation}
where the energy of zero oscillations stands on the right.

To calculate the quantity $\mathcal{UP}_{\mathbb{P}\mathbb{Q}}$
typical of the microtheory in the case of a quantum-mechanical
oscillator, i.e., in the state of equilibrium with a
quantum-mechanical thermostat, we use the expressions
\[
\Delta\mathbb{P}=\sqrt{\frac{\hbar
m\omega}{2}\coth\left(\varkappa\frac{\omega }{T_{0}}\right)},\]
\[\Delta\mathbb{Q}=\sqrt{\frac{\hbar}{2m\omega
}\coth\left(\varkappa\frac{\omega}{T_{0}}\right)},\] which are
contained in the wave functions of warm vacuum in the momentum and
coordinate representations, respectively. Then we obtain
\begin{equation}
\mathcal{U}\mathcal{P}_{\mathbb{P}\mathbb{Q}}=\frac{\hbar}{2}\coth
\left(\varkappa\frac{\omega}{T_{0}}\right).\label{51}%
\end{equation}
Comparing formulas (\ref{51}) and (\ref{36}), we obtain the SUR
\textquotedblleft coordinate--momentum\textquotedblright\ for a
quantum-mechanical oscillator in the state of equilibrium with the
quantum thermostat,
\begin{equation}
\mathcal{U}\mathcal{P}_{\mathbb{P}\mathbb{Q}}=\mathbb{J}_{0},\label{52}%
\end{equation}
with the minimum value $\mathbb{J}_{0}^{\mathrm{\min}}=\hbar/2$.

Note that, in contrast to SUR (\ref{49}), both the left- and
right-hand sides in SUR (\ref{52}) were derived in the framework of
the \textit{microtheory}. The saturated form of this SUR is
associated with the fact that the averaging of the corresponding
operators of stochastic action was carried out with the use of the
wave function depending on the temperature. The fact that both SURs
(\ref{49}) and (\ref{52}) are governed by the same macroparameter,
$\mathbb{J}_{0}$, is a substantial confirmation of the overlapping
of those theories, which are traditionally referred exclusively to
either micro- or macrodescriptions of the Nature.

It is worth demonstrating an interesting consequence of such an
overlapping. It is known that the restriction imposed on the value of
$\mathcal{UP}_{\mathbb{P}\mathbb{Q}}$ from below, which is equal to $\hbar/2$,
allows one to introduce, in the framework of the microtheory, elementary cells
$2\pi\hbar$ in dimensions in the \textquotedblleft
coordinate--momentum\textquotedblright\ phase space, which is a basis of
the quasiclassical approximation in the quantum-mechanical theory. Following the
analogy found above, it is also possible to introduce the \textquotedblleft
effective entropy--effective temperature\textquotedblright\ phase space in the
framework of the macrotheory. The restriction on $\mathcal{UP}_{\mathbb{S}%
\mathbb{T}}$ equal to $\frac{1}{2}\omega\hbar^{\ast}$ at an arbitrary
temperature means that this space is also partitioned into elementary cells of
$2\pi\omega\hbar^{\ast}$,\ the dimensions of which, however, increases with
the temperature.

In this connection, there emerges a tempting idea to treat the theory of
fluctuations of macroparameters in the framework of equilibrium stochastic
thermodynamics as a quasiclassical theory in the phase space of the variables
$(\mathbb{S},\mathbb{T})$. In its turn, the quasiclassical approximation in
the quantum-mechanical theory can be treated as a theory of
fluctuations of microparameters in the phase space of the variables $(p,q)$. Such an interpretation
allows the mathematical apparatus of the corresponding theories to be combined in
the framework of a synthetic theory, which would substantially cover both the
traditional thermodynamics and the traditional quantum-mechanical theory.

\section{Conclusions}

To summarize, we have proposed an approach, which allows the
main paradox of the standard theory of fluctuations of macroparameters to be
overcome. The paradox is associated with the fact that the conventional method
of taking the quantum-mechanical effects into account leads to that the results
obtained go beyond the scope of the thermodynamic description. As a result, on the
basis of the quantum-mechanical model of thermostat and the microdescription
within the $(\hbar,k)$-dynamics, a theory of quantum-thermal fluctuations
of effective macroparameters and their correlation has been developed, with
the corresponding thermodynamic description being preserved.

It has been demonstrated that the effective action, which is
regarded as a macroparameter that reflects the stochastic influence
of the environment, is responsible for the formation of
corresponding dispersions and correlators. We also found that the
uncertainties product  of conjugate macroparameters, the effective
entropy and the effective temperature, which characterizes the area
of an elementary cell in the phase space, is confined from below by
the energy of zero oscillations when approaching the absolute zero
temperature.

In addition, we have showed that the correlators of the pairs of conjugate effective
macroparameters $(\mathbb{S},\mathbb{T})$ and $(\mathbb{P},\mathbb{Q})$ in the
equilibrium state are proportional to each other at any Kelvin temperature and
linearly depend on the same macroparameter, the effective action
$\mathbb{J}_{0}$\ of the quantum thermostat.

At last, we have demonstrated that the minimum values of $\mathcal{UP}%
_{\mathbb{P}\mathbb{Q}}$ and $\mathcal{UP}_{\mathbb{S}\mathbb{T}}$ are
determined by the same world constant; it is Planck's constant. It allows the
theory of fluctuations of effective macroparameters to be regarded as a
quasiclassical theory in the phase space of the variables $(\mathbb{S}%
,\mathbb{T})$, and the quasiclassical theory in the phase space of
the variables $(p,q)$ as a theory of microparameters fluctuations.

\vskip3mm We consider as a pleasant duty to express our gratitude to
A.G.~Zagorodny and Yu.P.~Rybakov for their attention to the
organization of seminars at the M.M.~Bogoliubov Institute for
Theoretical Physics of the NAS of Ukraine (Kyiv) and at the
Department of theoretical physics of the Peoples' Friendship
University of Russia (Moscow), respectively, as well as the
participants at those seminars for the useful discussions. The work was
sponsored by the Russian Foundation for Basic Research (project
N~10-01-90408).

\rezume{%
КВАНТОВО-ТЕПЛОВІ ФЛУКТУАЦІЇ ЕФЕКТИВНИХ\\ МАКРОПАРАМЕТРІВ ТА ЇХ
КОРЕЛЯЦІЇ}{О.Д.  Суханов, О.Н. Голубєва, В.Г. Бар'яхтар} {Показано,
що використання стандартної теорії флуктуацій макропараметрів
удеяких випадках приводить до виходу за межі термодинамічного опису.
Сформульовано основи теорії квантово-теплових флуктуацій ефективних
макропараметрів та їх кореляції, узгодженої з умовами застосовності
рівноважної термодинаміки і заснованої на ефективних
макропараметрах, що враховують цілісний
стохастичний\rule{0pt}{9.5pt} вплив оточення за
будь-яких\rule{0pt}{9.5pt} температур. Обчислено корелятор спряжених
макропараметрів -- ефективної ентропії та
ефективної\rule{0pt}{9.5pt} температури -- і встановлено його
пропорційність ефективному впливу,\rule{0pt}{9.5pt} що характеризує
стохастичне оточення. Продемонстровано,\rule{0pt}{9.5pt} що
корелятори пар\rule{0pt}{9.5pt} спряжених ефективних параметрів
``ентропія--температура''\rule{0pt}{9.5pt} і ``координата--імпульс''
лінійно залежать від ефективного впливу,\rule{0pt}{9.5pt} а їх
мінімальні значення визначаються сталою Планка.\rule{0pt}{9.5pt}}

\end{document}